# Epitaxial metals for interconnects beyond Cu

Katayun Barmak [1,a)], Sameer Ezzat [2,3], Ryan Gusley [4], Atharv Jog [5], Sit Kerdsongpanya [5], Asim Khanya [6], Erik Milosevic [5], William Richardson [6], Kadir Sentosun [1], Amirali Zangiabadi [1], Daniel Gall [5], William E. Kaden [6], Eduardo R. Mucciolo [6], Patrick K. Schelling [6,7], Alan C. West [4], Kevin R. Coffey [8]

[1] Department of Applied Physics and Applied Mathematics, Columbia University, 500 West 120th Street, New York, NY 10027
[2] Department of Chemistry, University of Central Florida, 4111 Libra Drive, Orlando, FL 32816
[3] Department of Chemistry, University of Mosul, Mosul, Iraq
[4] Department of Chemical Engineering, Columbia University, 500 West 120th Street, New York, NY 10027
[5] Department of Materials Science and Engineering, Rensselaer Polytechnic Institute, 110 Eighth Street, Troy, NY 12180
[6] Department of Physics, University of Central Florida, 4111 Libra Drive, Orlando, FL 32816
[7] Advanced Materials Processing and Analysis Center, University of Central Florida, 4000 Central Florida Blvd. Box 162455, Orlando, FL 32816
[8] Department of Materials Science and Engineering and Department of Physics, University of Central Florida, 12760 Pegasus Drive, Orlando, FL 32816

a) Electronic mail: kb2612@columbia.edu

The experimentally measured resistivity of Co(0001) and Ru(0001) single crystal thin films, grown on c-plane sapphire substrates, as a function of thickness is modeled using the semiclassical model of Fuchs-Sondheimer. The model fits show that the resistivity of Ru would cross below that for Co at a thickness of approximately 20 nm. For Ru films with thicknesses above 20 nm, transmission electron microscopy evidences threading and misfit dislocations, stacking faults and deformation twins. Exposure of Co films to ambient air, and the deposition of oxide layers of $SiO_2$, MgO, $Al_2O_3$ and $Cr_2O_3$ on Ru degrade the surface specularity of the metallic layer. However, for the Ru films, annealing in a reducing ambient restores the surface specularity. Epitaxial electrochemical deposition of Co on epitaxially-deposited Ru layers is used as an example to demonstrate the feasibility of generating epitaxial interconnects for back-end of line structures.  An electron transport model based on a tight-binding (TB) approach is described, with Ru interconnects used an example. The model allows conductivity to be computed for structures comprising large ensembles of atoms ($10^5$-$10^6$), scales linearly with system size and can also incorporate defects.

## I.  INTRODUCTION

The continuous scaling of copper interconnects has produced two major challenges for CMOS technology – first, with respect to power and performance, and, second, with respect to reliability. Copper interconnects are the major source of power consumption in today's semiconductor devices and limit computing performance.[1,2] Furthermore, as the critical dimensions of interconnects are scaled towards, and then below the mean free path of Cu (39 nm at room temperature), the resistivity is found to increase.[2-7] The increase in resistivity, termed the resistivity size-effect, results in resistance scaling beyond Ohm's law dimensional scaling and leads to even larger power consumption and further limits improvements in computing performance.

The resistivity size-effect is typically attributed to the momentum loss of carriers along the axis of the conductor due to surface scattering (evidenced by the film-thickness, /line-width dependence of resistivity) and grain boundary scattering (evidenced by the grain size dependence of resistivity).  To date, the two most widely used physical models for these scattering mechanisms are the semiclassical models of Fuchs-Sondheimer (FS) for surface scattering and Mayadas-Shatzkes (MS) for grain boundary scattering.  The FS model is expressed in Eq. 1a and incorporates a specularity parameter, $p$, in the range of 0-1 for specular vs. diffuse scattering from surfaces.[8,9] The MS model is expressed in Eq. 1b and incorporates a reflection coefficient, $R$, in the range of 0-1 for scattering from grain boundaries.[10]

$$\rho_{FS} = \rho_o \left[ 1 - \left(\frac{3\lambda}{2d}\right)(1-p)\int_1^\infty \left(\frac{1}{t^3} - \frac{1}{t^5}\right)\frac{1-\exp(-kt)}{1-p\exp(-kt)}dt \right]^{-1} \quad (1a)$$

where $d$ is the film thickness and $\rho_o$ is the bulk resistivity, and

$$\rho_{MS} = \rho_o \left[ 1 - \frac{3}{2}\alpha + 3\alpha^2 - 3\alpha^3 \ln\left(1+\frac{1}{\alpha}\right) \right]^{-1} \quad (1b)$$

where $\alpha = \left(\lambda/g\right) R/(1-R)$ and $g$ is the grain size.

For both scattering mechanisms, the resistivity increase is seen to scale with the product of the bulk resistivity and the mean free path, $\rho_o \lambda$. Thus, a number of efforts directed towards identifying candidate metals for interconnects beyond Cu have focused on metals with lower values of $\rho_o \lambda$, as identified in Gall[11] based on DFT computed electron mean free path values.[11,12] Co and Ru are two of the metals that have lower values of $\rho_o \lambda$ compared to Cu and are the subject of the current report.

Furthermore, for polycrystalline interconnects where the grain size is of the order of $\lambda$, grain boundary scattering is expected to be the dominant scattering mechanism and has been shown to be the case for Cu films and lines.[3,5,13] Therefore, the development of interconnects beyond Cu points to the need for elimination of grain boundaries and the implementation of epitaxial metals as interconnects. This motivates the current work on experimental studies of the resistivity behavior of epitaxial, single crystal Co and Ru films.

In addition to experimental studies of epitaxial interconnects, there is a need for resistivity modeling approaches that go beyond the semiclassical models since for metals with non-spherical Fermi surfaces and even for Cu with its near-spherical Fermi surface at dimensions of the order of 10 nm and below, the semiclassical models have been shown to fail and crystallographic anisotropy of the size effect has been reported.[14-19] The models to be developed must also be able to describe electron transport for technologically relevant interconnect length scales and thus be able to model ensembles with large numbers of atoms (of the order of $10^5$-$10^6$). Furthermore, the newly developed models should allow incorporation of lattice vibrations and crystalline imperfections, including not only those within the volume of the interconnect, but also surface imperfections such as roughness and interfaces with dielectric encapsulants. In this work, we describe our efforts to date to develop a tight-binding (TB) model to describe electron transport in Ru lines with and without surface roughness, and to do so using an algorithm that scales linearly rather than with the cube of the number of atomic sites.

With respect to improved interconnect reliability, candidate metals should have higher melting points than Cu, with the expectation that this would lead to longer electromigration lifetimes on account of lower atomic mobility.[20-22] Co and Ru also meet this requirement with melting points respectively of 1495 and 2334 compared to 1057 °C for Cu. Since grain boundaries also act as preferred diffusion pathways for electromigration failure compared with diffusion through the bulk lattice, improved interconnect reliability again points to benefits of epitaxial, single crystal metals for interconnects beyond Cu.

However, if epitaxial, single crystal metal interconnects are to be implemented in CMOS technology, it is necessary to develop suitable metal deposition methods that can be integrated into back-end-of-line (BEOL) processing.[23] Electrochemical deposition is a room- or near room-temperature process and thus it is a natural candidate for satisfying the thermal budget constraints for BEOL. Furthermore, given the current use of electrodeposition for the fabrication of Cu interconnects, its use will extend the use of current process technologies and would not require the scale of investment that was necessary in transitioning from Al to Cu as the interconnect metal of choice.[24]  In this report, we also address capability of electrodeposition to fabricate epitaxial, single crystal metal films of Co. Additionally, a damascene-like process is proposed for the interconnect fabrication process and integration into BEOL.

The paper is structured as follows. Section II addresses the tight binding modeling approach to electron transport calculations of Ru lines. Section III reports the experimental resistivity behavior of epitaxial Ru and Co films, and the impact of oxides, whether formed by exposure to the ambient air, or by deposition of an oxide layer, and the effect of subsequent annealing treatments in oxidizing or reducing atmospheres on surface specularity and film resistivity.  Section IV presents electrochemical deposition of epitaxial Co on epitaxial layers of Ru.  Section V provides a summary and concludes the paper.

## II. MODELLING NANOWIRE CONDUCTION

Modeling of systems relevant to experimental results is a challenging problem. Specifically, approaches which can address transport in systems with perhaps $10^5$-$10^6$ atomic sites are required. This is generally beyond the realm of applicability of density-functional theory (DFT) using a plane-wave basis. Even using a limited set of basis states, the problem of directly solving the single-particle Schrödinger equation is impractical at these scales. Large-scale transport calculations require not only a limited set of basis states to define the single-particle Hamiltonian, but also an efficient algorithm that scales essentially linearly with the number of atom sites. Moreover, the objective should be to generate a realistic model which has the capability to describe scattering from surface roughness and thermal vibrations of the atomic sites.

The approach that we have taken is to apply a tight-binding (TB) model with an efficient algorithm to evaluate the electronic conductivity. We focus here on ruthenium nanowires described with the TB model developed in Ref. 25. In the TB approach, the single-particle eigenstates are expanded in a basis of atomic-like orbitals. To describe Ru metal, the 5s, 5p, and 5d orbitals are included in the model, for a total of 9 basis states per atom. However, in contrast to the model in Ref. 25, we have assumed that the basis states are orthonormal. To test the ability of this model to describe transport, we have computed the electron energy bands along high-symmetry directions in the Brillouin zone and compared the results to DFT computed bands using the VASP electronic structure code.[26-29] For the DFT calculations, we used pseudopotentials developed using the projector-augmented wave (PAW) method[30,31], and the Perdew-Wang[32] exchange and correlation functional was used. The results, especially for bands near the Fermi level, are shown in Fig. 1. Generally good agreement is found, suggesting that the model should be suitable for transport calculations. However, there are features that can be improved by refitting the model with the assumption of orthogonal basis states. We will return to some of the important considerations towards the end of this section.

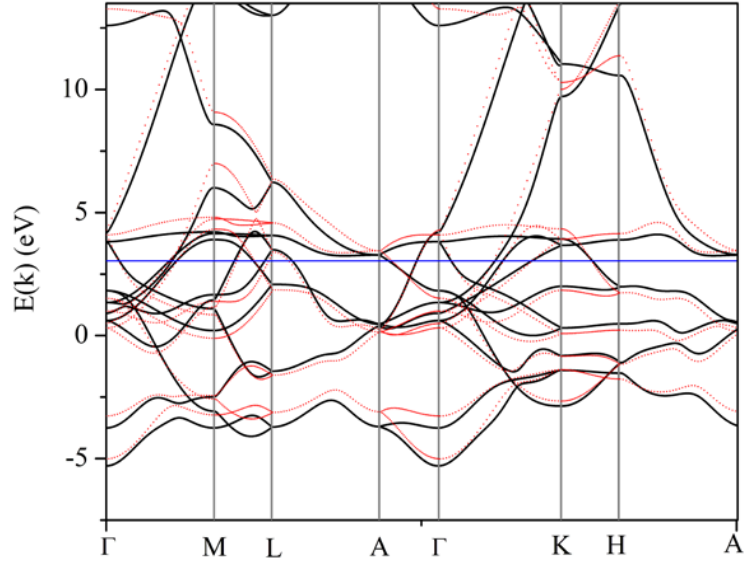

**Fig. 1** – Tight-binding (black solid lines) compared to DFT (red dashed lines) for the electronic structure of ruthenium in an hcp lattice. The comparison is made along high-symmetry directions within the Brillouin zone. The blue horizontal line indicates the Fermi energy.

The next important consideration is an efficient algorithm for evaluation of the electronic conductivity. The starting point is the Kubo-Greenwood equation for the conductivity,

$$\sigma_{xx}(E) = \frac{\pi \hbar e^2}{\Omega} Tr\left[ \hat{v}_x \, \delta(E - \hat{H}) \, \hat{v}_x \, \delta(E - \hat{H}) \right] \qquad (2)$$

in which $\sigma_{xx}(E)$ is the conductivity along the direction of the wire evaluated at energy $E$. The Hamiltonian $\hat{H}$ is evaluated using the TB basis states, with the hopping integrals between orbitals on different sites computed using the Slater-Koster integrals. Finally, the operator $\hat{v}_x$ is the velocity operator which is defined in the TB basis. When Eq. 2 is evaluated at the Fermi energy $E = E_F$, or alternately evaluated at several energies near the Fermi energy with the results convoluted using a Fermi-distribution function, the computed conductivity values should be directly comparable to experiment.

To efficiently evaluate Eq. 2, the kernel-polynomial method (KPM)[33] was used. The basic idea is to expand the Dirac-delta functions in Eq. 2 into a series of Chebyshev polynomials. The number of terms retained in the Chebyshev expansion directly determines the resolution of the Dirac delta functions in Eq. 2, which can be associated to a model-imposed inverse inelastic scattering time of the charge carriers (akin to phonon scattering). When static scattering is sufficiently weak (i.e., weak atomic disorder), this results in computed conductivity values that scale linearly with the number of retained terms. Finally, the trace in Eq. 2 is done using the truncated basis approximation, which employs a summation using a few random vectors. Because the Hamiltonian has a finite range for electron hopping, and only a few random vectors are required for the trace, the algorithm scales approximately linearly with system size.

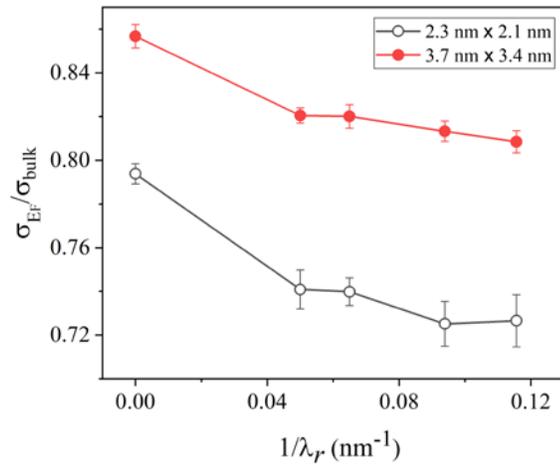

**Fig. 2** – Conductivity computed for wires with different cross-sectional areas and characteristic roughness length scales computed from Eq. 2 and the KPM method. The conductivity values are scaled relative to the prediction for the bulk conductivity to demonstrate the decreased values for the nanowires. Error bars were determined using the statistical variations obtained from using a limited number of random vectors and a finite ensemble of roughness realizations.

To best compare with experimental results, we have computed Eq. 2 using the Chebyshev polynomial expansion with a limited number of retained moments to approximately obtain the experimental room temperature conductivity of the bulk system for a perfect lattice. Specifically, for a calculation on a bulk system with 442,368 sites and 309 moments in the Chebyshev expansion, the conductivity $\sigma(E_F) = 1.55 \times 10^5$ $\Omega^{-1}$cm$^{-1}$ (or resistivity of 6.45 µΩcm) for transport along $[11\bar{2}0]$ or any other equivalent direction. The same approach was then applied to nanowires of different diameters and different characteristic length-scales $\lambda_r$ of surface roughness. The wires were oriented so that the transport direction is along the $[11\bar{2}0]$ crystallographic direction in the hcp lattice. In Fig. 2, results for the conductivity scaled to the bulk conductivity, $\sigma(E_F)/\sigma_{bulk}$, are shown for two different cross-sectional areas. Each point in Fig. 2 was computed using 8 random vectors to estimate the trace in Eq. 2. The conductivities were computed for wires with cross-sectional dimensions 2.3 nm×2.1 nm (51,200 atomic sites) and 3.7 nm×3.4 nm (131,072 sites). All wires were 138.5 nm in length with periodic-boundary conditions applied along the transport direction. Each calculation placed the sites at their perfect-lattice coordinates without relaxation or added displacements to model lattice vibrations. Hence the results in Fig. 2 represent an attempt to determine the effect of surface roughness and wire dimensions on electronic transport. The cases with $1/\lambda_r = 0$ correspond to a wire with no surface roughness. Even in the absence of surface roughness, the results indicate a substantial decrease in the conductivity for narrower wires. Specifically, the narrow wire had a conductivity approximately 0.795 of the bulk value, while the wire with slightly larger width and height had a value approximately 0.860 of the bulk value. The explanation for this dependency on diameter will be discussed later, but we have verified that larger wire dimensions do converge to the bulk conductivity value. Adding surface roughness does tend to result in a decrease in the conductivity, but surprisingly the effect is not dramatic. The computed values were obtained by randomly sampling surface roughness with a characteristic length scale $\lambda_r$ drawn from a Gaussian distribution. For roughened wires, each point in Fig. 2 represents the average of an ensemble of wires, specifically 25 realizations for wires with the small cross section, and 10 realizations for the wires with a larger cross section. Generally, the roughness extended to one or at most two atomic layers deep and was only added on the top $(0001)$ surface. The roughness length scales used are quite long in comparison to the expected inelastic mean-free path for electron-phonon scattering.

The effect of scattering from surface roughness in these calculations is surprisingly small. The most dramatic effect appears to be due to decreased wire dimensions even for wires without surface roughness. Roughness decreased the conductivity less than 10% from the value obtained for the bulk Ru crystal. If the scattering were strongly diffusive rather than specular, one would expect a much larger effect since the wire dimensions are significantly less than the bulk inelastic mean free path which has been estimated to be 6.59 nm in Ref. 11. Moreover, the reason for decreased conductivity for perfectly-smooth wires, where surface scattering is completely specular, needs to be understood.

To understand the reason for the behavior in Fig. 2, we explored the distribution of electrons within the wires when there was perfect crystalline order. Specifically, we exactly solved the single-particle Schrödinger equation in the tight-binding basis. By obtaining the expansion coefficients for the eigenstates, we are able to compute the effective electronic charge on each site in equilibrium. The results shown in Fig. 3 demonstrate that the model actually predicts a transfer of electrons from the surface into the core of the nanowire. This likely explains the dependence of the conductivity on wire diameter shown in Fig. 2. Specifically, transfer of electrons from one region to another may be responsible for blocking some conductance channels. In this case, it may be that some bands associated more strongly with surface sites do not cross the Fermi level and hence are emptied of electrons and do not contribute a conductance channel. Similarly, bands deep inside the wire have excess electrons, which indicates the existence of filled bands that do not cross the Fermi level. If this effect results in fewer conductance channels, this would result in a conductivity that scales with system size even in the absence of surface roughness.

However, the prediction in Fig. 3 is unexpected because a metal should approximately maintain local charge neutrality even when a surface is present. This highlights an important and perhaps somewhat unrecognized challenge of using TB models for transport calculations that must be addressed. One possible approach to assess local neutrality would be to describe site energies by including

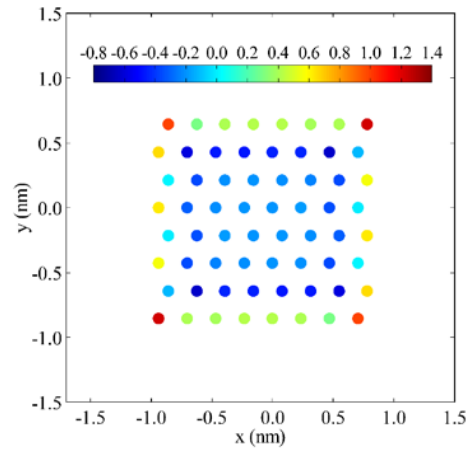

**Fig. 3** - End-on view of the charges obtained by exact solution of the single-particle Schrödinger equation for a perfect wire. The computed charges include the nuclear charge $Z = 8|e|$ for ruthenium. The results demonstrate that the TB model predicts a transfer of electrons from undercoordinated surface sites into bulk sites.

Coulomb interactions and ionization to the model within a self-consistent calculation. Models of this kind have been previously reported for metals including titanium.[34] The model in Ref. 34 was applied to surfaces, and it was found that charges on surface sites varied by no more than about $\pm 0.02|e|$. This is in contrast to the very large charging effects seen in the calculations in Fig. 3. However, the drawback of self-consistent calculations is that they cannot be used for very large systems where exact solutions of the Schrödinger equation become impractical. It might also be noted that the wires here were all terminated by an interface with vacuum rather than a dielectric material. Hence, although we expect that only minor charge rearrangement should occur at the metal-vacuum interface, significant charge rearrangement would be expected to occur for metal atoms bonding with an oxide dielectric. It would be expected that metal ions at an interface with a dielectric would become positively charged by donating electrons to form chemical bonds with oxygen ions. Modeling how this might affect transport and scattering at the interface could also be addressed by Coulomb self-constant models.

Instead, another approach that might be followed is to determine how the site energies used in the Hamiltonian should depend on their local environment such that local charge neutrality is maintained. This might be done, for example, using DFT calculations to generate a fitting database which includes not only bulk structures but also surfaces. Nanowires or nanoparticles might also be included in the fitting database. For an elemental metal like Ru, the condition of local neutrality might be assessed in the fitting process. This approach addresses an important uncertainty in the fitting of TB models, namely that the parameterization is usually determined by matching energy eigenvalues (i.e. the band structure) but without details related to the charge distribution. We suggest that going forward this should be taken into consideration especially when using TB models to predict transport properties that are specifically dependent on surface electron scattering.

We finally address the question of electron-phonon scattering. The model developed in Ref. 25 was fit to accurately predict the total energy and hence elastic properties. The approach used in the development of the model was to equate the total energy of the system to the sum of the energy eigenvalues up to the Fermi level. In Fig. 4, the computed cohesive energy as a function of volume is shown. Because our calculations require an orthogonal basis which is in contrast to the original model[25], the equilibrium volume is smaller than experimental values and the cohesive energy is somewhat larger. Specifically, the cohesive energy from experiment is approximately 6.62 eV at the equilibrium volume 13.61 Å$^3$.[25] The predictions from the TB model are clearly in strong disagreement based on the data plotted in Fig. 4. Before the model can be used for predictions of electron-phonon scattering, significant improvements in the model are necessary.

In summary, the TB approach using KPM to compute transport coefficients in nanowires is promising from the perspective of computational efficiency, but it is also clear that significant care needs to be exercised in determining the tight-binding parameterization. Specifically, models of elemental metals should assess the condition of local neutrality before use in transport calculations. It should also be possible to more directly model electron-phonon scattering with improvements in the ability of the TB model to compute the total energy as well as energetics associated with small atomic displacements. Models of this kind have been previously reported.[34] With improvements over the existing models, large-scale calculations will be possible that can access realistic wire dimensions for comparison to experiment, as well as detailed scattering predictions from surface roughness and lattice vibrations.

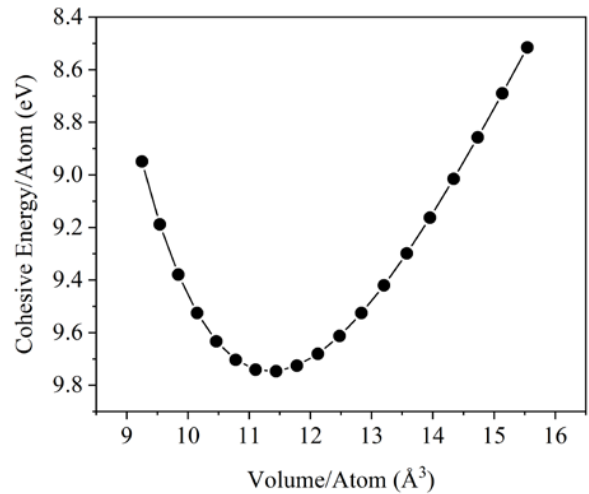

**Fig. 4** – Cohesive energy for ruthenium predicted by the TB model as a function of system volume per atom.

# III. EXPERIMENTAL STUDIES OF FILMS AND DIELECTRIC INTERFACES

## A. Co and Ru Films

The details of the preparation of the epitaxial Co and Ru films whose resistivities as a function of thickness are presented in Fig. 5 are reported elsewhere.[36-38] Briefly, the metal layers were deposited on $Al_2O_3$(0001) by DC magnetron sputtering.[39,40] The Co films were deposited at 300 °C and the Ru films at 350 °C. Immediately after deposition, the samples were subjected to *in situ* vacuum annealing in the same vacuum chamber. The annealing procedure consisted of a single-step anneal at 500 °C for 1 h for the Co layers and a temperature ramp with six consecutive 30-minute intervals at 450, 550, 650, 750, 850 and 950 °C for the Ru layers. The samples were cooled in the vacuum chamber for 12 h and were then transferred to an attached analysis chamber for *in situ* transport measurements without air exposure. The resistivity was measured using a four-point probe.[41,42] Subsequently, the samples were removed from the vacuum system via a load lock chamber that was vented with dry $N_2$, and immediately (within 2 s) submerged in liquid nitrogen to minimize air exposure prior to low temperature transport measurements. The resistivity at 77 K was then measured with both sample and measurement tips immersed in liquid nitrogen. *Ex situ* room-temperature resistivity measurements were taken after warming the samples up to 295 K using a continuous flux of $N_2$ gas to minimize water condensation on the sample surface. Resistivity measurements were also taken after 24-48 h, to confirm that extending the air-exposure beyond the initial approximately 2 minutes has a negligible effect on electron scattering.

Figure 5 is a plot of the Co(0001) and Ru(0001) resistivity $\rho$ vs film thickness $d$, measured from epitaxial layers both in vacuum (*in situ*) and air (*ex situ*) at 295 K, and immersed in liquid $N_2$ at 77 K. The resistivity increases with decreasing $d$ for all data sets, which is due to electron surface scattering. The Ru resistivities measured *in situ* and *ex situ* (solid and open diamonds in Fig. 5) are identical within experimental uncertainty, indicating that air exposure has a negligible effect on electron scattering at the Ru(0001) surface. In contrast, the *ex situ* values are larger than the *in situ* resistivity for Co(0001) (open vs. solid squares in Fig. 5), suggesting a decrease in the surface scattering specularity. The plotted resistivity at 77 K is considerably lower than at 295 K for both Ru and Co, due to the lower electron-phonon scattering at low temperatures. However, the resistivity increase due to surface scattering at low $d$ has a comparable magnitude at 295 and 77 K, indicating that (to first order approximation) electron surface scattering is temperature-independent and Matthiessen's rule applies.

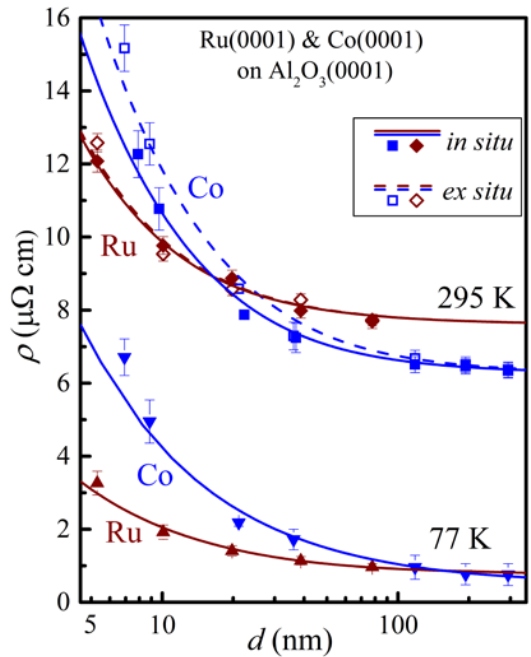

**Fig. 5** – Resistivity $\rho$ of epitaxial Ru(0001)/Al$_2$O$_3$(0001) (dark red diamonds) and Co(0001)/Al$_2$O$_3$(0001) (blue squares) films vs thickness $d$, measured *in situ* (solid symbols) and *ex situ* (open symbols) in vacuum and air at 295 K, and immersed in liquid $N_2$ at 77 K (triangles). Curves are from data fitting using the FS model (Eq. 1a). The plotted data is reorganized based on our previous published studies.[36-38,43]

We quantify the resistivity size effect by interpreting the measured data within the classical Fuchs-Sondheimer (FS) model with non-equal scattering specularities at the top and bottom surfaces,[4] as indicated by the solid and dashed lines in Fig. 5 which are the results from curve fitting. A well-known challenge when applying this FS model to describe resistivity vs. thickness data is that it does not allow for independent determination of the two parameters that quantify electron scattering, which are the surface scattering specularity $p$ and the bulk mean free path $\lambda$. More specifically, for any arbitrary choice of $p$, a value for $\lambda$ can be found such that the model predicts a $\rho$ vs $d$ curve that matches the measured data.[17] Therefore, as a first step, data fitting is done assuming completely diffuse electron scattering at both the upper and lower film surfaces ($p_u = p_l = 0$), yielding $\lambda$ values which can be interpreted as a lower bound for the mean free path. This yields $\lambda = 6.7 \pm 0.3$ nm for Ru measured both *in situ* and *ex situ*, and $\lambda = 14.0 \pm 0.5$ and $19.5 \pm 1.0$ nm for *in situ* and *ex situ* data points for Co, respectively. The corresponding curves in Fig. 5 are solid for *in situ* and dashed for *ex situ* measurements. The values at 77 K are $\lambda = 36.7 \pm 2.1$ nm for Ru and $\lambda = 217 \pm 20$ nm for Co. The increase in resistivity during air exposure of Co results in an apparent increase of $\lambda$, which conversely can be attributed to a decrease in the scattering specularity during surface oxidation, as we have previously quantified for Co in Ref. 37. More specifically, both *in situ* and *ex situ* data from Co can be simultaneously described with a $\lambda = 19.5$ nm and a scattering specularity of the top surface that decreases from $p_u = 0.55$ for the *in situ* measurements to $p_u = 0$ for *ex situ*, based on the assumption that Co surface oxidation results in completely diffuse electron surface scattering.

The direct comparison of the resistivity of Co and Ru in Fig. 5 shows a resistivity cross-over. More specifically, Co has a 21% lower bulk resistivity than Ru,[10] but $\rho_{Ru} < \rho_{Co}$ for thicknesses $d < 20$ nm, where 20 nm corresponds to the cross-over thickness where the resistivity of epitaxial Ru and Co layers is identical. The cross-over thickness is slightly larger ($d = 25$ nm) if considering the *ex situ* data,[43] which gives Ru an additional conductance benefit because of its inert surface. The figure also shows that the cross-over at 77 K is at a much larger thickness of $d = 160$ nm. This is because bulk scattering is smaller at low temperatures such that the size-dependent resistivity contribution dominates the overall resistivity at larger dimensions. Similarly, the resistivity benefit of Ru over Co becomes more dominant if considering lines instead of thin films, with an estimated resistivity cross-over at a 50-nm half-pitch for polycrystalline Ru vs Co lines.[36,43] This is due to four vs. two scattering surfaces for lines vs. thin films and the contribution from grain boundary scattering in polycrystalline lines (Eq. 1b).

## B. *Defects in Ru Films*

A subset of the Ru films of Fig. 5 were characterized by transmission electron microscopy (TEM). Two sets of cross sections were prepared by focused ion beam milling with the directions normal to the sections parallel to $[10\bar{1}0]$ and $[11\bar{2}0]$ directions of the Ru layer, respectively. Additional details of sample preparation and imaging are given elsewhere.[37,38] Figures 6a-c show high resolution TEM images of the nominally 10, 40, and 80 nm thick films. (The thicknesses of these films as determined by X-ray reflectivity are given in Refs. 37 and 38.) Presence of defects at an angle to the basal plane in the nominally 80 nm thick film in Fig. 6c, one of which is marked with an arrow, is clearly seen. The basal plane is parallel to the Ru/sapphire interface in the image.

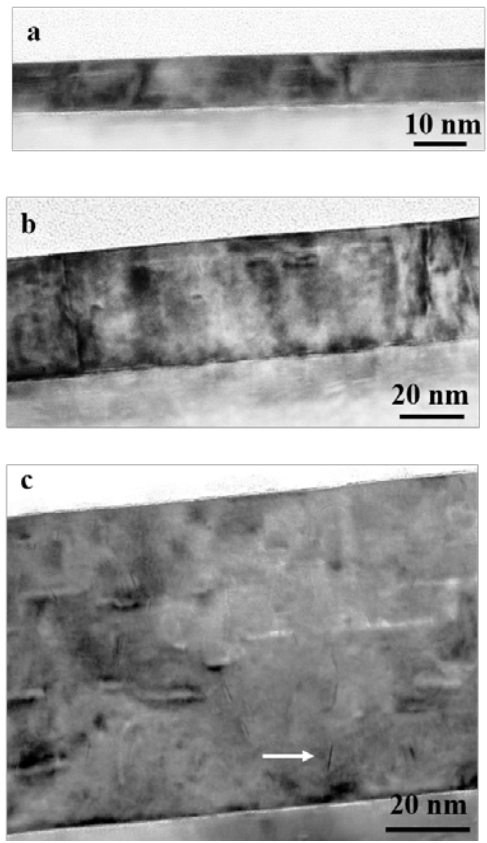

As the thickness of an epitaxially deposited film increases, the strain energy resulting from the misfit strain increases, and above a critical thickness, the strain relaxes by the formation of defects. This relaxation commonly occurs via slip i.e., via the formation of threading screw dislocations that glide to the interface and deposit misfit dislocations at the interface that relax the strain.[44-46] Deformation twinning is an alternative plastic deformation mode to slip. During twinning, the original (parent) lattice is re-oriented by atom displacements which are equivalent to a simple shear of the lattice points, or some integral fraction of these points.[47-49] The invariant plane of shear is termed $\mathbf{K}_1$, and the shear direction, or the twinning direction, is $\boldsymbol{\eta}_1$. Classical theory of twinning identifies four twinning systems, i.e., $\mathbf{K}_1$ $\boldsymbol{\eta}_1$ pairs.[47-49]

**Fig. 6** – High resolution transmission electron micrographs of the cross section of the epitaxial Ru films with nominal thicknesses of (a) 10, (b) 40, (c) 80 nm deposited on c-plane sapphire, imaged close to the Ru $\left[ 10\bar{1}0 \right]$ zone axis. The white arrow marks a deformation twin.

Defects in the Ru layers such as the one marked in Fig. 6c are deformation twins. To determine which of four twin systems is the one observed in the Ru films, the angle between the trace of the twinning planes in the image plane and the c-axis of the Ru layer, which is normal to the Ru/sapphire interface, was measured for both cross sections of the 80 nm-thick film. The trace of defect is the intersection line of defect plane and the image plane. Experimentally measured values were $14.7° \pm 2.1°$ for the Ru $(10\bar{1}0)$ image plane. This value is closest to the crystallography computed value of 17.5° (next closest was 32.3°). Therefore, we conclude that the defects at an angle to the basal plane are the $\{11\bar{2}1\}\frac{1}{3}\langle\bar{1}\bar{1}2\bar{6}\rangle$. (We note that the computed value of the angle is outside of the standard deviation of the experimental values, but this may well be because the sample is not exactly on zone axis for Ru, though it is exactly on zone axis for sapphire.)

In addition to deformation twins, Figs. 6b and c show the presence of a number of defects parallel to the basal plane. These defects are more clearly visualized in the weak beam dark field image presented in Fig. 7, which also evidences the presence of threading and misfit dislocations, the latter at the Ru/sapphire interface.[47-50] The defects parallel to the basal plane are stacking faults, which are also seen in epitaxially deposited AlN films with the Wurtzite structure deposited on sapphire substrates.[50] Weak beam dark field imaging and "**g.b**" analysis with three different **g**-vectors, i.e., reciprocal lattice vectors, were used to determine the Burgers vector, **b**, of the threading dislocations.[51] These dislocations were shown to be either **c**-type or **a**-type, i.e., with Burgers vectors of $\langle 0001 \rangle$ or $\langle 11\bar{2}0 \rangle$ type. The misfit dislocations are also either pure edge **a**-type or have an **a**-type edge component.[44]

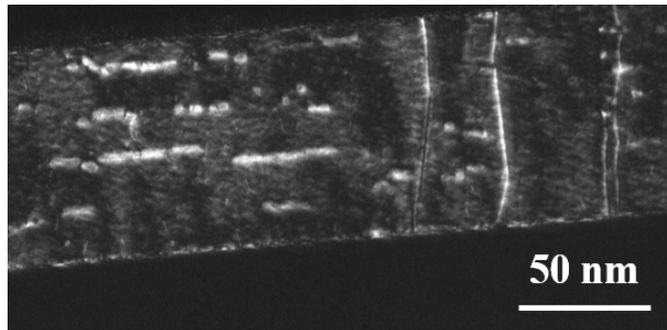

**Fig. 7** – Weak beam dark field transmission electron micrograph of the 80 nm-thick Ru film shown in Fig. 6.

The 40 nm-thick Ru film in Fig.6b and the 20 nm-thick Ru film (not shown in Fig. 6) exhibit similar types of defects to the 80 nm-thick film, but the density of defects appears to be lower. Significantly, the 10 nm-thick film shown in Fig. 6a exhibits no defects.

It is interesting to note that the apparently high density of defects seen in the nominally 80 nm-thick Ru films has no measurable impact on film resistivity, as evidenced by measured resistivity values that are equal to the in **c**-plane bulk resistivity of Ru within experimental error. The *in situ* measured resistivity of this film is $7.69 \pm 0.1$ μΩcm for a thickness measured by X-ray reflectivity of 77.8 nm.[38] The reported in-plane bulk resistivity of Ru is 7.6 μΩcm.[52,53]

### C.  Ru Films with Deposited Surface Oxides

For the Co films of Fig. 5, the degradation of surface specularity upon exposure to the ambient air points to the need for development of approaches enabling the controllable optimization of surface specularity to increase conductivity in nm-scale interconnects. To this end, our recent work[54] provides evidence for process-controlled optimization of these effects at $SiO_x$/Ru interfaces, briefly summarized below as motivation for investigation of the impact of other oxides reported in the current study.

In Ref. 54, several key properties expected to correlate with film resistivity were measured and compared to changes in modeled values for upper surface specularity inferred from measured values of film thickness and conductivity at each of 11 sequential processing steps.[54] In this example, a ~20 nm thick Ru film was first deposited via DC magnetron sputtering onto a 2-inch diameter c-axis sapphire, (0001)$Al_2O_3$, wafer maintained at 700 °C within a 4 mTorr Ar environment. After deposition, the wafer was removed from the sputter deposition vacuum chamber and cut into several ~7×7 $mm^2$ coupons for further processing and characterization. The sequential processing of a group of five coupons followed film deposition as step (1) and consisted of: (2) *ex situ* step-annealing to 950 °C within a flow of Ar/$H_2$ 3% maintained at 1 atm, (3) sputter deposition of a ~5 nm $SiO_x$ overcoat at 4 mTorr Ar at room temperature, (4) *ex situ* annealing to 500 °C within the Ar/$H_2$ environment, (5) annealing to 500 °C within 1 atm of air, (6) again annealing to 500 °C within the Ar/$H_2$ environment, (7) repetition of the reductive annealing conditions employed in step 2, (8) annealing to 400 °C within 1 atm Ar/$O_2$ 20%, (9) repetition of the reductive annealing conditions employed in steps 4 and 6, (10) annealing to 350 °C within 1 atm Ar/$O_2$ 20%, and (11) one last high temperature annealing cycle equivalent to those used in steps 2 and 7. At each step in the sequential process, samples were characterized using a battery of: (i) Van der Pauw 4-point probe measurements of sheet resistance, (ii) angle-dependent measurements of X-ray reflectivity (XRR), which were fit using simulated $SiO_2$/Ru(0001)/$Al_2O_3$(0001) layered models to establish layer thickness and interface roughness, and (iii) X-ray photoelectron spectroscopy (XPS), used to monitor variations in the relative abundance of interfacial Ru in oxidized and reduced chemical-states.

Between steps 4 and 11, the specularity of the upper Ru surface, as determined from a fit to the FS model, was found to inversely correlate with the extent of Ru oxidation measured by XPS, with changes to both properties appearing to be reversible by subsequent annealing within chemically reducing/oxidizing environments. Samples exhibiting increased Ru-oxidation by XPS simultaneously exhibit increased Ru upper surface roughness and increased $RuO_x$ with deceased Ru thicknesses by XRR. A similar improvement in specularity and decrease in roughness was also observed when annealing the as-deposited Ru film to 950 °C during the transition from step 1 to step 2, which coincided with an apparent improvement to the long-range order of the Ru interface as evidenced by sharpened low energy electron diffraction (LEED) patterns measured *ex situ* following the second step (relative to the first). Interestingly depositing an $SiO_x$ layer onto the pre-annealed Ru film (step 3) caused decreases in upper surface specularity equivalent to the largest fluctuations observed throughout the entirety of the full dataset, but without any of the other characteristic changes to surface properties noted in the other steps (i.e. Ru thickness, roughness, and oxidation did not appear to change within the uncertainty of the measured values when depositing $SiO_x$). Moreover, annealing the $SiO_x$-capped sample within $Ar/H_2$ fully restored Ru surface specularity to its uncovered value despite again causing no significant changes to any of the other tracked physical properties associated with the film.

The above results were taken to constitute a clear demonstration for an approach to controllably tune scattering behavior at metal/dielectric interfaces, which then stimulated our interest in exploring the effect of similar processing conditions on the conductivity of Ru films capped with different dielectrics. To this end, Figs. 8a and b provide an abridged set of analogous experiments to those in Ref. 54 conducted to compare changes in upper surface specularity following deposition and subsequent annealing of $SiO_2$, MgO, $Al_2O_3$, and $Cr_2O_3$ capped Ru films. The initial processing step shown here is the same as the 2nd step listed above (i.e., Ru deposition and subsequent step-annealing to 950 °C in Ar/$H_2$, defined as process step 1 in Fig. 8).[54] As expected for the equivalent processing up to this point (no oxide deposition), similar resistivities are observed for all four groups of samples. The next process step for the four groups of samples was room temperature sputter deposition of ~5 nm thick overlayers of the indicated dielectric materials, and this results in a significant increase in film resistivity, which is again interpreted as increased surface scattering induced by the presence of the oxides. As shown, this resistivity increase is noticeably greater for films overcoated with MgO. Next, the coupons were annealed at 500°C in Ar/H$_2$, which decreases the film resistivities to values comparable to those noted after the first process step (i.e. before deposition of the oxides), and further annealing of the films within Ar/H$_2$ at elevated temperatures (950 °C) results in even lower resistivities. In general, these changes in resistivity are similar to those discussed above for SiO$_2$/Ru interfaces. Of interest is the greater range of resistivity observed for the samples overcoated with MgO, which will require further study to fully understand.

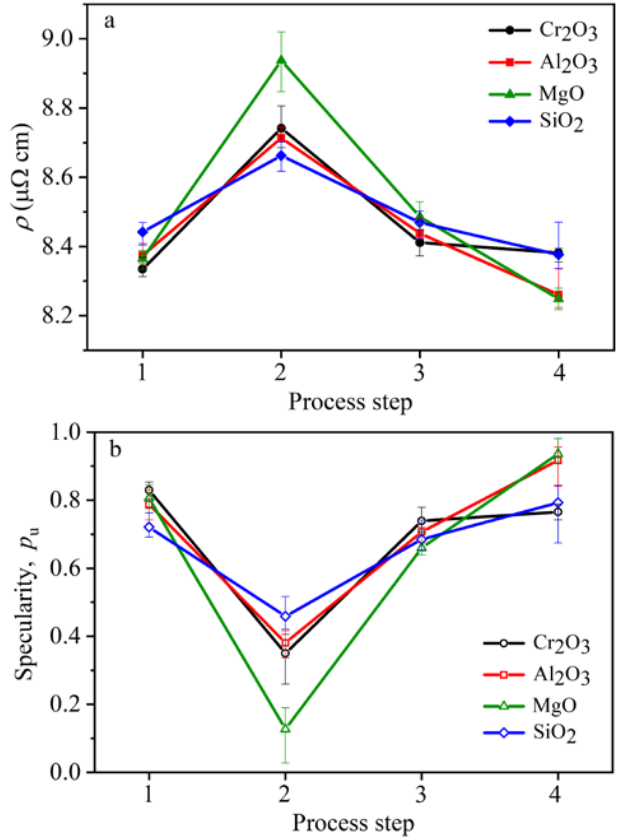

**Fig. 8** – Variation of (a) resistivity and (b) upper surface specularity for sequentially processed groups of five ~20-nm thick deposited oxide/Ru/sapphire samples. The plotted points are the average values for the five samples while the error bars indicate the highest and lowest values amongst the five. Process steps are as follows: (1) Annealing at 950 °C in Ar+$H_2$ following film deposition (2) deposition of oxide layer, (3) annealing at 500 °C in Ar+$H_2$ and (4) annealing at 950 °C in Ar+$H_2$.

The changes in resistivity noted in Fig. 8a are understood to be due to changes in the Ru upper surface (Ru/dielectric or Ru/air interface) as the lower Ru surface (Ru/sapphire interface) and the bulk of the Ru layer had previously been annealed to 950°C and are presumed to be unchanged by room temperature dielectric overlayer deposition and subsequent 500°C and 950°C anneals (an assumption also applied previously[54]). As was noted earlier, the values chosen for $\lambda$ and $p$ within the FS model are not unique, as from inspection of Eq. 1a it is clear that only the product, $\lambda(1-p)$, is significant. However, the average specularity, $p$, as well as the upper and lower surface specularity values for the lower and upper surfaces, $p_l$ and $p_u$, respectively, are constrained by the model to be in the range of $p = 0$ (fully diffuse scattering) to $p = 1$ (fully specular scattering). To satisfy these boundary conditions, the larger range of resistivities observed for the samples with MgO/Ru interfaces require changes to the $\lambda$ and $p_l$ values chosen relative to those used in the prior report[54]. Whereas upper surface Ru specularities $p_u$ calculated in the prior report[54] assumed $\lambda = 11.0$ nm, and $p_l = 0.0$, the same parameters must be changed to 12.5 nm and 0.30, respectively, to produce physically meaningful results in Fig. 8b, which reflect full-scale (0-1) fluctuation in upper Ru surface specularity as a function of MgO coverage and annealing conditions.

In summary, the formation of surface oxides upon exposure to ambient air in the case of Co, and deposition of oxide layers on epitaxial metal films and annealing treatments in oxidizing gases at elevated temperatures in the case of Ru, show degradation of surface specularity of the metal layers. Combined with the subsequent restoration of the Ru surface specularity upon annealing treatment in a reducing atmosphere, it is clear that surface specularity can be controllably varied. However, the need to use different values of $\lambda$ and $p$ to fit the FS model for a given resistivity-thickness data set clearly points to the shortcomings of this semiclassical model in describing electron transport in metals in the deep nanoscale regime. This then underscores the need for alternative models, with attempts to date for one such model described in Section II.

## IV. ELECTRODEPOSITION OF EPITAXIAL METAL FILMS

As noted in the Introduction, implementation of epitaxial, single crystal metal interconnects in CMOS technology requires metal deposition methods that can be integrated into back-end-of-line (BEOL) processing, with electrodeposition as the most likely candidate method. To this end, in recent work and in the current report, the ability to use electrodeposition for epitaxial deposition of Co layers is demonstrated.

The details of electrodeposition of the Co layers are given elsewhere.[55] Briefly, the electrolyte consisted of 1 mM cobalt sulfate heptahydrate, 0.125 mM sulfuric acid, 10 mM potassium sulfate and 0.1 mM potassium chloride, with the pH of the solution 3.8.

The electrodeposition was carried out using a three-electrode set-up with Pt as the counter electrode and Ag/AgCl as the reference electrode. The substrates for the electrodeposition were 10 and 60 nm-thick Ru(0001) layers sputter deposited onto c-plane sapphire, the former at 400 °C and the latter at 500 °C. Following deposition, the films were *ex situ* step annealed from 450 to 950 °C in 100 °C steps and held at each temperature for 30 minutes. Prior to electrodeposition of the Co layer, the surface oxides from Ru were reduced using a potentiostatic hold in 50 mM sulfuric acid, and the electrodeposition bath was purged with Ar for one hour to remove dissolved oxygen.

The Nernst potential of Co deposition from the solution was calculated as -0.57 V. The open circuit potential after deposition of Co was measured as -0.58 V, in close agreement with the calculated value. In Ref. 55, cyclic voltammetry (CV) and linear sweep voltammetry (LSV), the latter from various hold potentials, were used to demonstrate the underpotential deposition (UPD) of both Co and hydrogen at potentials above their respective reversible potentials. UPD is a self-limiting process that results in the deposition of (sub)monolayer of the depositing species at a potential positive of bulk deposition, or overpotential deposition (OPD).

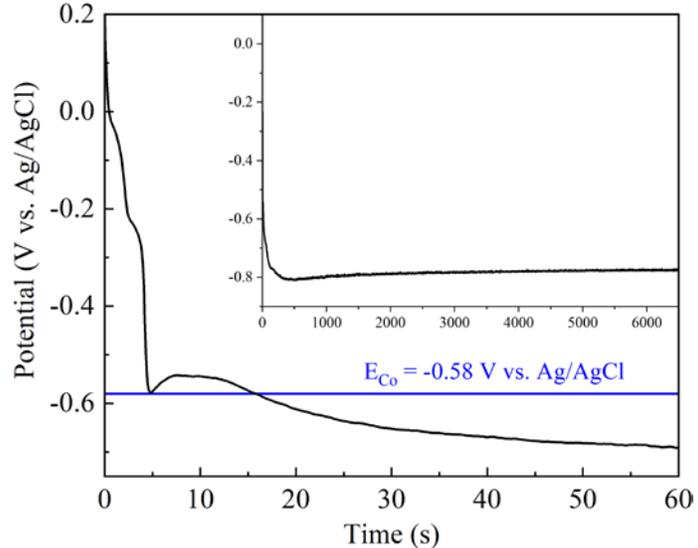

**Fig. 9** – Potential transient for Co electrodeposition on a 60 nm-thick Ru(0001) film at -80 µA/cm$^2$. At short times, the potential resides above $E_{Co}$ as Co is first deposited underpotentially as a (sub)monolayer. At long times (see inset), the potential remains constant at -0.78 V vs. Ag/AgCl.

UPD of a metal layer occurs when it is more energetically favorable for it to deposit onto a foreign substrate than it is for that metal to deposit onto itself.

To generate epitaxial, single crystal Co layers, electrodeposition was conducted at constant current rather than constant potential. Figure 9 shows the potential transient resulting from the electrodeposition of Co at -80 µA/cm$^2$ onto the 60 nm-thick Ru layer. At short times during the galvanostatic deposition, the potential reaches the standard reduction potential of Co before returning to potential values of 10-40 mV above the reduction potential. This implies that Co is first deposited underpotentially as a (sub)monolayer. After about 16 seconds, the potential reaches values negative of E$_{Co}$ which allows for the deposition of epitaxial Co overlayers as the overpotential deposition (OPD) reaction proceeds. At long times, the potential remains constant around -0.78 V. Deposition for 7200 seconds at -80 µA/cm$^2$ results in an epitaxial Co film that has an average thickness of 25 nm.

X-ray diffraction (XRD) and transmission electron microscopy were used to characterize the resulting Co film and to show that the Co layer was indeed epitaxial, single crystal with a hexagon-on-hexagon orientation relationship to the Ru layer.[55] The hexagon-on-hexagon orientation relationship of Co to the Ru layer should be contrasted with the rotated honeycomb orientation relationship of Ru to sapphire.[37,56]

The orientation relationship and the single crystal, epitaxial nature of the Co layer is demonstrated here in Figs. 10a and b for a Co layer deposited at -80 µA/cm$^2$ for 7200 s onto a 10 nm-thick Ru layer. Figure 10a is the selected area electron diffraction pattern for a cross-sectional sample of the Co/Ru/sapphire stack with the electron beam along the $[10\bar{1}0]$ zone axis of the Ru layer and thus the $[11\bar{2}0]$ zone axis of the sapphire substrate. For the pairs of diffraction spots seen Fig. 10a, the inner spots are from both Ru and sapphire since a 10 nm Ru layer, unlike the 60 nm Ru layer, is epitaxially strained and lattice matched to the sapphire. The outer diffraction spots are from the Co layer, which clearly evidence the relaxation of strain in the Co layer.

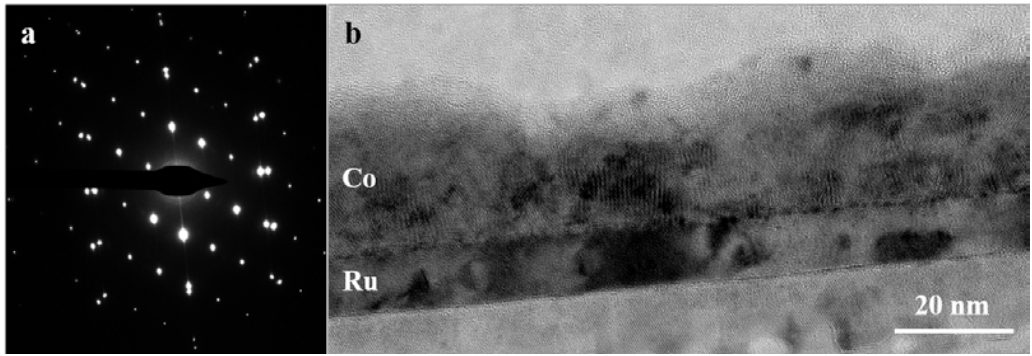

**Fig. 10** – Cross section of a Co layer electrodeposited onto a 10 nm-thick sputter deposited epitaxial Ru layer on c-plane sapphire. (a) Selected area diffraction pattern, and (b) high resolution transmission electron micrograph. The zone axis for Ru and Co is $[10\bar{1}0]$ while for sapphire it is $[11\bar{2}0]$ on account of the orientation relationships between the metal layers and the substrate. See text for more detail.

The presence of both Co and the Ru layers is seen in the high-resolution transmission electron micrograph of Fig. 10b. Additional higher magnification high resolution transmission electron microscopy (not shown) clearly demonstrates the epitaxial growth of Co on Ru with parallel (0002) layers close to the Co/Ru interface.[55] Figure 10b also shows that whereas the Ru layer is planar, the Co layer is rough. We expect that the control of the surface roughness of the Co layer will require modifications to the electrodeposition procedure to minimize the impact of hydrogen reduction that occurs simultaneously with Co reduction. It will also require the use of epitaxial underlayers that are better lattice matched to Co than Ru is. However, Figs. 10a and b and Ref. 55 clearly provide a proof-of-principle demonstration of epitaxial electrodeposition to be used for interconnects beyond Cu.

In prior work[55], it was noted that in order to implement epitaxial electrodeposition into the interconnect fabrication process, it is necessary to have access to a single crystal, conductive seed layer such as a silicide layer formed epitaxially on the single crystal base semiconductor wafer.[57] Intermediate epitaxial layers may be required to improve the lattice matching of the silicide layer to the interconnect metal of interest. The electrodeposited epitaxial interconnect metal will then be grown from the bottom of the vias upwards to the next metallization layer. Continued lateral growth from these vias until impingement of growth from adjacent vias will allow trenches to be filled with the single crystal interconnect metal connecting devices, in an essentially analogous manner to the damascene process currently in use for Cu interconnects. Subsequent planarization can then define each layer of the interconnect metallization.

## IV. SUMMARY AND CONCLUSIONS

Resistivity as a function of thickness was presented for epitaxial Co(0001) and Ru(0001) films grown on c-plane sapphire substrates. It was shown that within the context of the semiclassical FS surface scattering model, the resistivity of Ru would cross below that for Co at a thickness of approximately 20 nm. The defects in the epitaxial Ru films were characterized by transmission electron microscopy. For thicknesses above 20 nm, threading and misfit dislocations, stacking faults and deformation twins were found to be present. The 10 nm thick films were free of these defects.

Formation of an oxide layer upon exposure of the Co films to ambient air, and the deposition of oxide layers of $SiO_2$, $MgO$, $Al_2O_3$ and $Cr_2O_3$ on Ru were shown to degrade the surface specularity of the metallic layer. However, for the Ru films, annealing in a reducing ambient restored the surface specularity, and for the case of silicon dioxide repeated annealing in oxidizing and then reducing ambients was shown to decrease and then again restore the surface specularity.

When using the FS model, the need to vary the values of surface specularity and the electron mean free path to fit different resistivity data sets for a given metal pointed to the need for development of new transport models. Efforts to date to develop models based on the tight-binding (TB) approach to electron transport that scale linearly with system size and allow conductivity to computed for interconnects with $10^5$-$10^6$ atomic sites and incorporate bulk and surface defects were described. Future TB-based efforts require model improvements that enforce local charge neutrality.

Epitaxial electrochemical deposition of Co on epitaxially-deposited Ru layers on c-plane sapphire substrates was used as an example to demonstrate the feasibility of epitaxial deposition at temperatures compatible with CMOS processing. A damascene like approach similar to that currently used for Cu, but taking advantage of epitaxial underlayers such as contact silicides for epitaxial electrodeposition, was proposed as a method for implementing epitaxial metals for interconnects beyond Cu.

## ACKNOWLEDGMENTS

Funding support by the National Science Foundation grants ECCS-1740228, 1740270 and 1740271, and the E2CDA-NRI Program of the Semiconductor Research Corporation under Tasks 2764.001 and 2764.003, as well as support from the Air Force Office of Scientific Research AFOSR FA9550-18-1-0063 and FA9550-19-1-0156 is gratefully acknowledged. This work was carried out in part in the Electron Microscopy Laboratory

**Figure Captions**

**Fig. 1** – Tight-binding (black solid lines) compared to DFT (red dashed lines) for the electronic structure of ruthenium in an hcp lattice. The comparison is made along high-symmetry directions within the Brillouin zone. The blue horizontal line indicates the Fermi energy.

**Fig. 2** – Conductivity computed for wires with different cross-sectional areas and characteristic roughness length scales computed from Eq. 2 and the KPM method. The conductivity values are scaled relative to the prediction for the bulk conductivity to demonstrate the decreased values for the nanowires. Error bars were determined using the statistical variations obtained from using a limited number of random vectors and a finite ensemble of roughness realizations.

**Fig. 3** - End-on view of the charges obtained by exact solution of the single-particle Schrödinger equation for a perfect wire. The computed charges include the nuclear charge $Z = 8|e|$ for ruthenium. The results demonstrate that the TB model predicts a transfer of electrons from undercoordinated surface sites into bulk sites.

**Fig. 4** – Cohesive energy for ruthenium predicted by the TB model as a function of system volume per atom.

**Fig. 5** – Resistivity $\rho$ of epitaxial Ru(0001)/Al$_2$O$_3$(0001) (dark red diamonds) and Co(0001)/Al$_2$O$_3$(0001) (blue squares) films vs thickness $d$, measured *in situ* (solid

symbols) and *ex situ* (open symbols) in vacuum and air at 295 K, and immersed in liquid $N_2$ at 77 K (triangles). Curves are from data fitting using the FS model (Eq. 1a). The plotted data is reorganized based on our previous published studies.[36-38,43]

**Fig. 6** – High resolution transmission electron micrographs of the cross section of the epitaxial Ru films with nominal thicknesses of (a) 10, (b) 40, (c) 80 nm deposited on c-plane sapphire, imaged close to the Ru $[10\bar{1}0]$ zone axis. The white arrow marks a deformation twin.

**Fig. 7** – Weak beam dark field transmission electron micrograph of the 80 nm-thick Ru film shown in Fig. 6.

**Fig. 8** – Variation of (a) resistivity and (b) upper surface specularity for sequentially processed groups of five ~20- nm thick deposited oxide/Ru/sapphire samples. The plotted points are the average values for the five samples while the error bars indicate the highest and lowest values amongst the five. Process steps are as follows: (1) Annealing at 950 °C in Ar+$H_2$ following film deposition (2) deposition of oxide layer, (3) annealing at 500 °C in Ar+$H_2$ and (4) annealing at 950 °C in Ar+$H_2$.

**Fig. 9** – Potential transient for Co electrodeposition on a 60 nm-thick Ru(0001) film at -80 µA/cm$^2$. At short times, the potential resides above $E_{Co}$ as Co is first deposited underpotentially as a (sub)monolayer. At long times (see inset), the potential remains constant at -0.78 V vs. Ag/AgCl.

**Fig. 10** – Cross section of a Co layer electrodeposited onto a 10 nm-thick sputter deposited epitaxial Ru layer on c-plane sapphire. (a) Selected area diffraction pattern, and (b) high resolution transmission electron micrograph. The zone axis for Ru and Co is $[10\bar{1}0]$ while for sapphire it is $[11\bar{2}0]$ on account of the orientation relationships between the metal layers and the substrate. See text for more detail.